\documentclass[10pt,a4paper]{article}
\usepackage{lineno}
\usepackage{graphicx}
\usepackage{xspace}
\usepackage{color}
\usepackage{colortbl}
\usepackage{subfigure}
\usepackage{multirow}
\usepackage{amsmath}
\usepackage{ifthen}
\newboolean{articletitles}
\setboolean{articletitles}{true} 
\newboolean{uprightparticles}
\setboolean{uprightparticles}{false}

\usepackage{rotating} 
\usepackage{amssymb}
\usepackage{amsfonts}
\usepackage{upgreek}
\usepackage{hyperref}
\usepackage[all]{hypcap} 
\usepackage{bm}
\usepackage{afterpage}
\usepackage{mciteplus}

\def\lhcb {LHCb\xspace}
\def\ux85 {UX85\xspace}

\def\babar  {BaBar\xspace}
\def\belle  {Belle\xspace}

\def\ads {{\rm ADS}\xspace}

\ifthenelse{\boolean{uprightparticles}}%
{
 
 \def\Pgamma      {\ensuremath{\upgamma}\xspace}

 \def\Ppi         {\ensuremath{\uppi}\xspace}

 \def\Ppsi        {\ensuremath{\uppsi}\xspace}

 \def\PDelta      {\ensuremath{\Delta}\xspace}                 
 \def\PXi      {\ensuremath{\Xi}\xspace}                 
 \def\PLambda      {\ensuremath{\Lambda}\xspace}                 
 \def\PSigma      {\ensuremath{\Sigma}\xspace}                 
 \def\POmega      {\ensuremath{\Omega}\xspace}                 
 \def\PUpsilon      {\ensuremath{\Upsilon}\xspace}                 
 
 \def\PB      {\ensuremath{\mathrm{B}}\xspace}                 
                  
 \def\PD      {\ensuremath{\mathrm{D}}\xspace}

 \def\PJ      {\ensuremath{\mathrm{J}}\xspace}                 
 \def\PK      {\ensuremath{\mathrm{K}}\xspace}

 \def\Pb      {\ensuremath{\mathrm{b}}\xspace}

 \def\Pi      {\ensuremath{\mathrm{i}}\xspace}

 \def\Ps      {\ensuremath{\mathrm{s}}\xspace}

}
{
 
 \def\Pgamma      {\ensuremath{\gamma}\xspace}

 \def\Ppi         {\ensuremath{\pi}\xspace}

 \def\Ppsi        {\ensuremath{\psi}\xspace}                 
                  
 \mathchardef\PDelta="7101
 \mathchardef\PXi="7104
 \mathchardef\PLambda="7103
 \mathchardef\PSigma="7106
 \mathchardef\POmega="710A
 \mathchardef\PUpsilon="7107
                  
 \def\PB      {\ensuremath{B}\xspace}                 
                  
 \def\PD      {\ensuremath{D}\xspace}

 \def\PJ      {\ensuremath{J}\xspace}                 
 \def\PK      {\ensuremath{K}\xspace}

 \def\Pb      {\ensuremath{b}\xspace}

 \def\Pi      {\ensuremath{i}\xspace}

 \def\Ps      {\ensuremath{s}\xspace}

}

\def\g      {\ensuremath{\Pgamma}\xspace}

\def\squark    {\ensuremath{\Ps}\xspace}

\def\bquark    {\ensuremath{\Pb}\xspace}
\def\bquarkbar {\ensuremath{\overline \bquark}\xspace}
\def\bbbar     {\ensuremath{\bquark\bquarkbar}\xspace}

\def\pion  {\ensuremath{\Ppi}\xspace}

\def\pip   {\ensuremath{\pion^+}\xspace}
\def\pim   {\ensuremath{\pion^-}\xspace}

\def\pipm  {\ensuremath{\pion^\pm}\xspace}
\def\pimp  {\ensuremath{\pion^\mp}\xspace}

\def\kaon  {\ensuremath{\PK}\xspace}
  \def\Kbar  {\kern 0.2em\overline{\kern -0.2em \PK}{}\xspace}

\def\Kz    {\ensuremath{\kaon^0}\xspace}
\def\Kzb   {\ensuremath{\Kbar^0}\xspace}
\def\KzKzb {\ensuremath{\Kz \kern -0.16em \Kzb}\xspace}
\def\Kp    {\ensuremath{\kaon^+}\xspace}
\def\Km    {\ensuremath{\kaon^-}\xspace}
\def\Kpm   {\ensuremath{\kaon^\pm}\xspace}
\def\Kmp   {\ensuremath{\kaon^\mp}\xspace}
\def\KpKm  {\ensuremath{\Kp \kern -0.16em \Km}\xspace}
\def\KS    {\ensuremath{\kaon^0_{\rm\scriptscriptstyle S}}\xspace}

  \def\Dbar    {\kern 0.2em\overline{\kern -0.2em \PD}{}\xspace}
\def\D       {\ensuremath{\PD}\xspace}

\def\Dz      {\ensuremath{\D^0}\xspace}
\def\Dzb     {\ensuremath{\Dbar^0}\xspace}
\def\DzDzb   {\ensuremath{\Dz {\kern -0.16em \Dzb}}\xspace}
\def\Dp      {\ensuremath{\D^+}\xspace}
\def\Dm      {\ensuremath{\D^-}\xspace}

\def\DpDm    {\ensuremath{\Dp {\kern -0.16em \Dm}}\xspace}

\def\Dstarpm {\ensuremath{\D^{*\pm}}\xspace}

\def\B       {\ensuremath{\PB}\xspace}
  \def\Bbar    {\kern 0.18em\overline{\kern -0.18em \PB}{}\xspace}

\def\Bz      {\ensuremath{\B^0}\xspace}

\def\Bu      {\ensuremath{\B^+}\xspace}
\def\Bub     {\ensuremath{\B^-}\xspace}
\def\Bp      {\ensuremath{\Bu}\xspace}
\def\Bm      {\ensuremath{\Bub}\xspace}
\def\Bpm     {\ensuremath{\B^\pm}\xspace}

\def\Bs      {\ensuremath{\B^0_\squark}\xspace}
\def\Bsb     {\ensuremath{\Bbar^0_\squark}\xspace}

\def\jpsi     {\ensuremath{{\PJ\mskip -3mu/\mskip -2mu\Ppsi\mskip 2mu}}\xspace}
\def\psitwos  {\ensuremath{\Ppsi{(2S)}}\xspace}

  \def\Y#1S{\ensuremath{\PUpsilon{(#1S)}}\xspace}% no space before {...}!

\def\ra                 {\ensuremath{\rightarrow}\xspace}
\def\to                 {\ensuremath{\rightarrow}\xspace}

\def\CP                {\ensuremath{C\!P}\xspace}

\def\AT#1     {\ensuremath{A_{\mathrm{T}}^{#1}}\xspace}           % 2

\def\C#1      {\ensuremath{\mathcal{C}_{#1}}\xspace}                       % 9
\def\Cp#1     {\ensuremath{\mathcal{C}_{#1}^{'}}\xspace}                    % 7
\def\Ceff#1   {\ensuremath{\mathcal{C}_{#1}^{\mathrm{(eff)}}}\xspace}        % 9  
\def\Cpeff#1  {\ensuremath{\mathcal{C}_{#1}^{'\mathrm{(eff)}}}\xspace}       % 7
\def\Ope#1    {\ensuremath{\mathcal{O}_{#1}}\xspace}                       % 2
\def\Opep#1   {\ensuremath{\mathcal{O}_{#1}^{'}}\xspace}                    % 7

\newcommand{\tev}{\ensuremath{\mathrm{\,Te\kern -0.1em V}}\xspace}
\newcommand{\gev}{\ensuremath{\mathrm{\,Ge\kern -0.1em V}}\xspace}
\newcommand{\mev}{\ensuremath{\mathrm{\,Me\kern -0.1em V}}\xspace}
\newcommand{\kev}{\ensuremath{\mathrm{\,ke\kern -0.1em V}}\xspace}
\newcommand{\ev}{\ensuremath{\mathrm{\,e\kern -0.1em V}}\xspace}
\newcommand{\gevc}{\ensuremath{{\mathrm{\,Ge\kern -0.1em V\!/}c}}\xspace}
\newcommand{\mevc}{\ensuremath{{\mathrm{\,Me\kern -0.1em V\!/}c}}\xspace}
\newcommand{\gevcc}{\ensuremath{{\mathrm{\,Ge\kern -0.1em V\!/}c^2}}\xspace}
\newcommand{\gevgevcccc}{\ensuremath{{\mathrm{\,Ge\kern -0.1em V^2\!/}c^4}}\xspace}
\newcommand{\mevcc}{\ensuremath{{\mathrm{\,Me\kern -0.1em V\!/}c^2}}\xspace}

\def\invpb {\ensuremath{\mbox{\,pb}^{-1}}\xspace}

\def\invfb   {\ensuremath{\mbox{\,fb}^{-1}}\xspace}

\def\gsim{{~\raise.15em\hbox{$>$}\kern-.85em
          \lower.35em\hbox{$\sim$}~}\xspace}
\def\lsim{{~\raise.15em\hbox{$<$}\kern-.85em
          \lower.35em\hbox{$\sim$}~}\xspace}

\def\pt         {\mbox{$p_{\rm T}$}\xspace}

\def\evtgen     {\mbox{\textsc{EvtGen}}\xspace}
\def\pythia     {\mbox{\textsc{Pythia}}\xspace}

\def\geant      {\mbox{\textsc{Geant4}}\xspace}

\def\tell1  {TELL1\xspace}
\def\ukl1   {UKL1\xspace}

\newcommand{\eg}{\mbox{\itshape e.g.}\xspace}

\newcommand{\bea}{\begin{eqnarray}}
\newcommand{\eea}{\end{eqnarray}}
\newcommand{\beq}{\begin{equation}}
\newcommand{\eeq}{\end{equation}}

\begin{document}
\begin{titlepage}
\belowpdfbookmark{Title page}{title}
\pagenumbering{roman}
\vspace*{-1.5cm}
\centerline{\large EUROPEAN ORGANIZATION FOR NUCLEAR RESEARCH (CERN)}
\vspace*{1.5cm}
\hspace*{-5mm}\begin{tabular*}{\textwidth}{lc@{\extracolsep{\fill}}r}
\vspace*{-13mm}\mbox{\!\!\!\includegraphics[width=.12\textwidth]{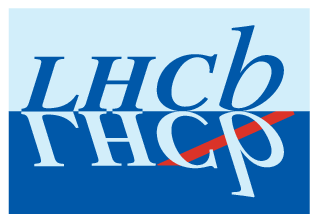}}& & \\
&& LHCb-PAPER-2012-001\\
&& CERN-PH-EP-2012-071\\
&&\today \\
\end{tabular*}
\vspace*{4cm}
\begin{center}
{\bf\Large\boldmath 
Observation of $C\!P$ violation in $B^{\pm}\to DK^{\pm}$ decays
}\\
\vspace{2cm}
\normalsize { The LHCb collaboration }
\footnote{Authors are listed on the following pages.}
\vspace{2cm}
\begin{abstract}
  \noindent
An analysis of $B^{\pm}\to DK^{\pm}$ and $B^{\pm}\to D\pi^{\pm}$ decays is presented where the $D$ meson is reconstructed in the two-body final states: $K^{\pm}\pi^{\mp}$, $K^+K^-$, $\pi^+\pi^-$ and $\pi^{\pm}K^{\mp}$.
Using $1.0{\rm \,fb}^{-1}$ of LHCb data, measurements of several observables are made including the first observation of the suppressed mode $B^{\pm}\to [\pi^{\pm}K^{\mp}]_DK^{\pm}$.
$C\!P$ violation in $B^{\pm}\to DK^{\pm}$ decays is observed with $5.8\,\sigma$ significance.
\end{abstract}
\vspace{2cm}
{\it Submitted to Physics Letters B}\\
\vspace{4cm}
\end{center}
\end{titlepage}

\onecolumn
\begin{center}
\begin{small}
{\bf \lhcb collaboration}\\
\begin{flushleft}
%{\Large LHCb Collaboration ----- official authorship list}\\[4ex]
%valid for date: 6. Feb. 2012\\
%used for paper: Observation of CP violation in $B^\pm$ decays (LHCb-PAPER-2012-001)\\[4ex]
%collaborators included, who did not leave before 6. Feb. 2011\\
%                           and who joined before 6. Aug. 2011\\[2ex]
%{\small today is 2. Feb. 2012}\\[4ex]
%-- 
%-- LHCb Authorlist, Status of 6. Feb. 2012
%-- 
R.~Aaij$^{38}$, 
C.~Abellan~Beteta$^{33,n}$, 
B.~Adeva$^{34}$, 
M.~Adinolfi$^{43}$, 
C.~Adrover$^{6}$, 
A.~Affolder$^{49}$, 
Z.~Ajaltouni$^{5}$, 
J.~Albrecht$^{35}$, 
F.~Alessio$^{35}$, 
M.~Alexander$^{48}$
% ...

{\footnotesize \it
% ...
$ ^{5}$Clermont Universit\'{e}, Universit\'{e} Blaise Pascal, CNRS/IN2P3, LPC, Clermont-Ferrand, France\\
$ ^{6}$CPPM, Aix-Marseille Universit\'{e}, CNRS/IN2P3, Marseille, France\\
% ...
$ ^{33}$Universitat de Barcelona, Barcelona, Spain\\
$ ^{34}$Universidad de Santiago de Compostela, Santiago de Compostela, Spain\\
$ ^{35}$European Organization for Nuclear Research (CERN), Geneva, Switzerland\\
% ...
$ ^{38}$Nikhef National Institute for Subatomic Physics, Amsterdam, The Netherlands\\
% ...
$ ^{43}$H.H. Wills Physics Laboratory, University of Bristol, Bristol, United Kingdom\\
% ...
$ ^{48}$School of Physics and Astronomy, University of Glasgow, Glasgow, United Kingdom\\
$ ^{49}$Oliver Lodge Laboratory, University of Liverpool, Liverpool, United Kingdom\\
% ...
$ ^{n}$LIFAELS, La Salle, Universitat Ramon Llull, Barcelona, Spain\\
% ...
}
\bigskip
%---- LHCb Authorlist, Status 6. Feb. 2012
%---- Number of Authors = 591
%---- 
\end{flushleft}
\end{small}
\end{center}
\newpage
\twocolumn
\pagestyle{empty} 
\pagestyle{plain}
\setcounter{page}{1}
\pagenumbering{arabic}

\section{Introduction}
A fundamental feature of the Standard Model and its three quark generations is that all \CP violation phenomena are the result of a single phase in the CKM quark-mixing matrix~\cite{Cabibbo:1963yz,*Kobayashi:1973fv}.
The validity of this model may be tested in several ways, and one --- verifying the unitarity condition $V_{ud}V_{ub}^*+V_{cd}V_{cb}^*+V_{td}V_{tb}^*=0$ --- is readily applicable to $B$ mesons.
This condition describes a triangle in the complex plane whose area is proportional to the amount of \CP violation in the model~\cite{PhysRevLett.55.1039}.
Following the observation of \CP violation in the \Bz system~\cite{Aubert:2001nu,*Abe:2001xe}, the focus has turned to testing the unitarity of the theory by over-constraining the sides and angles of this triangle.
Most related measurements involve loop or box diagrams, and for which the CKM model is typically assumed when interpreting data~\cite{PhysRevD.84.033005,*Bona:2005vz}.
This means the least-well determined observable, the phase $\g=\arg\left(-V_{ud}V_{ub}^* / V_{cd}V_{cb}^* \right)$ is of particular interest as $\g\neq0$ can produce direct \CP violation in tree decays.\\

One of the most powerful methods for determining \g is measurements of the partial widths of $\Bpm\to D\Kpm$ decays where the $D$ signifies a \Dz or \Dzb meson.
In this case, the amplitude for the $\Bm\to\Dz\Km$ contribution is proportional to $V_{cb}$ whilst the $\Bm\to\Dzb\Km$ amplitude depends on $V_{ub}$.
If the $D$ final state is accessible for both \Dz and \Dzb mesons, the interference of these two processes gives sensitivity to \g and may exhibit direct \CP violation.
This feature of open-charm \Bm decays was first recognised in its application to \CP eigenstates, 
such as $D\to\Kp\Km$, $\pip\pim$~\cite{Gronau:1990ra,*Gronau:1991dp} but can be extended to other decays, \eg\ $D\to\pim\Kp$.
This second category, labelled ``ADS" modes in reference to the authors of~\cite{Atwood:1996ci,*Atwood:2000ck}, requires the favoured, $b\to c$ decay to be followed by a doubly Cabibbo-suppressed $D$ decay, 
and the suppressed $b\to u$ decay to precede a favoured $D$ decay. 
The amplitudes of such combinations are of similar total magnitude and hence large interference can occur.
For both the \CP-mode and \ads methods, the interesting observables are partial widths and \CP asymmetries.\\

In this paper, we present measurements of the \Bpm decays in the \CP modes, $[\Kp\Km]_Dh^{\pm}$ and $[\pip\pim]_Dh^{\pm}$, 
the suppressed \ads mode $[\pipm K^{\mp}]_Dh^{\pm}$ and the favoured $[\Kpm\pi^{\mp}]_Dh^{\pm}$ combination where $h$ indicates either pion or kaon.
Decays where the bachelor  --- the charged hadron from  the \Bm decay --- is a kaon carry greater sensitivity to \g. 
$\Bm\to D\pim$ decays have some limited sensitivity and provide a high-statistics control sample from which probability density functions (PDFs) are shaped.
In total, 13 observables are measured: three ratios of partial widths
\beq
R_{K/\pi}^f = \frac{ \Gamma(\Bm\to [f]_D\Km)+\Gamma(\Bp\to [f]_D\Kp) }{ \Gamma(\Bm\to [f]_D\pim)+\Gamma(\Bp\to [f]_D\pip) },
\eeq
\noindent
where $f$ represents $KK$, $\pi\pi$ and the favoured $K\pi$ mode, six \CP asymmetries
\beq
A_{h}^f = \frac{ \Gamma(\Bm\to [f]_Dh^-)-\Gamma(\Bp\to [f]_Dh^+) }{ \Gamma(\Bm\to [f]_Dh^-)+\Gamma(\Bp\to [f]_Dh^+) },
\eeq
\noindent
and four charge-separated partial widths of the \ads mode relative to the favoured mode
\beq
R_h^{\pm}  = \frac{ \Gamma(\Bpm\to [\pipm\Kmp]_Dh^{\pm})}{ \Gamma(\Bpm\to  [\Kpm\pimp]_Dh^{\pm})}.
\eeq
Elsewhere, similar analyses have established the $\Bpm\to D_{\CP}h^{\pm}$ modes
\cite{delAmoSanchez:2010ji,Abe:2006hc,Aaltonen:2009hz}
and found evidence of the $\Bpm\to [\pipm\Kmp]_D\Kpm$ decay~\cite{Belle:2011ac,delAmoSanchez:2010dz,Aaltonen:2011uu}.
Analyses of $\Bpm\to [\KS h^+h^-]_D\Kpm$ decays~\cite{delAmoSanchez:2010rq,Poluektov:2010wz}
have yielded the most precise measurements of \g though a $5\sigma$ observation of \CP violation from a single analysis has not been achieved.
This work represents the first simultaneous analysis of $\Bpm\to D_{\CP}h^{\pm}$ and $\Bpm\to D_{\!\ads}h^{\pm}$ modes.
It is motivated by the future extraction of \g which, with this combination, may be determined with minimal ambiguity.\\

This paper describes an analysis of 1.0~\invfb of $\sqrt{s}=7~\tev$ data collected by \lhcb in 2011.
The 2010 sample of 35~\invpb is used to define the selection criteria in an unbiased manner.
The \lhcb experiment~\cite{Alves:2008zz} takes advantage of the high $b\bar{b}$ and $c\bar{c}$ cross sections at the Large Hadron Collider to record large samples of heavy hadron decays.
It instruments the pseudorapidity range $2<\eta <5$ of the proton-proton ($pp$) collisions with a dipole magnet and a tracking system which achieves a momentum resolution of $0.4-0.6\%$ in the range $5-100$~\gevc. 
The dipole magnet can be operated in either polarity and this feature is used to reduce systematic effects due to detector asymmetries. In 2011, 58\% of data were taken with one polarity, 42\% with the other.
The $pp$ collisions take place inside a silicon microstrip vertex detector that provides clear separation of secondary $B$ vertices from the primary collision vertex (PV) as well as discrimination for tertiary $D$ vertices.
Two ring-imaging Cherenkov (RICH) detectors with three radiators (aerogel, $C_{4}F_{10}$ and $CF_4$) provide dedicated particle identification (PID) which is critical for the separation of $\Bm\! \to\! D \Km$ and $\Bm\! \to\! D \pim$ decays.\\

A two-stage trigger is employed. 
First a hardware-based decision is taken at a frequency up to 40~MHz. 
It accepts high transverse energy clusters in either an electromagnetic calorimeter or hadron calorimeter, or a muon of high transverse momentum (\pt).
For this analysis, it is required that one of the three tracks forming the \Bpm candidate points at a deposit in the hadron calorimeter, or that the hardware-trigger decision was taken independently of these tracks.
A second trigger level, implemented entirely in software, receives 1 MHz of events and retains $\sim0.3\%$ of them.
It searches for a track with large \pt and large impact parameter (IP) with respect to the PV.
This track is then required to be part of a secondary vertex with a high \pt sum, significantly displaced from the PV.
In order to maximise efficiency at an acceptable trigger rate, the displaced vertex is selected with a decision tree algorithm that uses \pt, $\chi^2_{\rm IP}$, flight distance and track separation information. 
Full event reconstruction occurs offline, and after preselection around $2.5\times10^{5}$ events are available for final analysis.\\

Approximately one million simulated events for each $B^\pm \to[h^+h^-]_D h^\pm$ signal mode are used as well as a large inclusive sample of generic $B\to DX$ decays.
These samples are generated using a tuned version of \pythia~\cite{Sjostrand:2006za} to model the $pp$ collisions,
\evtgen~\cite{Lange:2001uf} encodes the particle decays and \geant~\cite{Agostinelli:2002hh} to describe interactions in the detector. 
Although the shapes of the signal peaks are determined directly on data, the inclusive sample assists in the understanding of the background.
The signal samples are used to estimate the relative efficiency in the detection of modes that differ only by the bachelor track flavour.

\section{Event selection}
During event reconstruction, 16 combinations of $B^\pm \to D h^\pm$, $D \to h^\pm h^\mp$ are formed with the candidate $D$ mass within $1765-1965$~\mevcc.
$D$ daughter tracks are required to have $\pt>250$~\mevc but this requirement is tightened to $0.5<\pt<10$~\gevc and $5<p<100$~\gevc for bachelor tracks to ensure best pion versus kaon discrimination.
The decay chain is refitted~\cite{2005NIMPA.552..566H} constraining the vertices to points in space and the $D$ candidate to its nominal mass, $m^{D^0}_{\rm PDG}$~\cite{Nakamura:2010zzi}.\\

Reconstructed candidates are selected using a boosted decision tree (BDT) discriminator~\cite{Hocker:2007ht}.
It is trained using a simulated sample of $\Bpm \to [K^\pm \pi^\mp]_D\Kpm$ and background events from the $D$ sideband ($35<|m(hh)-m^{D^0}_{\rm PDG}|<100$~\mevcc) of the independent sample collected in 2010. The BDT uses the following properties of the candidate \Bpm decay:
\begin{itemize}
\item From the tracks, the $D$ and \Bpm:  \pt and $\chi^2_{\rm IP}$ with respect to the PV;
\item From the \Bpm and $D$:  decay time, flight distance from the PV and vertex quality;
\item From the \Bpm: the angle between the momentum vector and a line connecting the PV to its decay vertex.
\end{itemize}
Information from the rest of the event is employed via an isolation variable that considers the imbalance of \pt around the \Bpm candidate,
\begin{equation}
A_{\pt}=\frac{\pt(B)-\sum_n\pt}{\pt(B)+\sum_n\pt},
\end{equation}
where the $\sum_n\pt$ sums over the $n$ tracks within a cone around the candidate excluding the three signal tracks.
The cone is defined by a circle of radius 1.5 in the plane of pseudorapidity and azimuthal angle (measured in radians).
As no PID information is used as part of the BDT, it performs equally well for all modes considered here.\\

The optimal cut value on the BDT response is chosen by considering the combinatorial background level ($b$) in the invariant mass distribution of favoured $\Bpm\to[K\pi]_D\pipm$ candidates.
The large signal peak in this sample is scaled to the anticipated \ads-mode branching fraction to provide a signal estimate ($s$). The quantity $s/\sqrt{s + b}$ serves as an optimisation metric.
The BDT response peaks towards 0 for background and 1 for signal.
The optimal cut is found to be $>0.92$ for the \ads mode; this is also applied to the favoured mode.
For the cleaner \CP modes, a cut of ${\rm BDT}>0.80$ gives a similar background level but with a 20\% higher signal efficiency.\\

PID information is quantified as differences between the logarithm of likelihoods, $\ln\mathcal{L}_h$, under five mass hypotheses, $h\in\{\pi,K,p,e,\mu\}$ (DLL).
Daughter kaons of the $D$ meson are required to have ${\rm DLL}_{K\pi}=\ln\mathcal{L}_{K}-\ln\mathcal{L}_{\pi}>2$ and daughter pion must have ${\rm DLL}_{K\pi}<-2$.
Multiple candidates are arbitrated by choosing the candidate with the best-quality \Bpm vertex; only 26 events in the final sample of $157\, 927$ require this consideration.\\

Candidates from $B$ decays that do not contain a true $D$ meson can be reduced by requiring the flight distance significance of the $D$ candidate from the \Bm vertex to be $>2$.
The effectiveness of this cut is monitored in the $D$ sideband where it is seen to remove significant structures peaking near the \Bm mass.
A simulation study of the $\Bm\to\Km\Kp\Km$, $\Km\pip\pim$ and $\Km\Kp\pim$ modes suggests this cut leaves 2.5, 1.3 and 0.8 events respectively under the $\Bm\to[\Kp\Km]_D\Km$, $[\pip\pim]_D\Km$ and $[\pip\Km]_D\Km$ signals.
This cut also removes cross feed ({\it e.g.} $\Bm\to[\Km\pip]_D\pim$ as a background of $[\pip\pim]_D\Km$) which occurs when the bachelor is confused with a $D$ daughter at low decay time.
Finally, the combination of the bachelor and the opposite-sign \Dz daughter is made under the hypothesis they are muons. 
The parent \B candidate is vetoed if the invariant mass of this combination is within $\pm22$~\mevcc of either the \jpsi or \psitwos mass~\cite{Nakamura:2010zzi}.\\

Due to misalignment, the reconstructed \Bpm mass is not identical to the established value, $m^{\Bpm}_{\rm PDG}$ \cite{Nakamura:2010zzi}. 
As simulation is used to define background shapes, it is useful to apply linear momentum scaling factors separately to the two polarity datasets so the \Bpm mass peak is closer to $m^{\Bpm}_{\rm PDG}$.
After this correction, the $\Dz\to\Km\pip$ mass peak is measured at 1864.8~\mevcc with a resolution of 7.4~\mevcc.
Selected $D$ candidates are required to be within $\pm25$~\mevcc of $m^{D^0}_{\rm PDG}$. This cut is tight enough that no cross feed occurs from the favoured mode into the \CP modes.
In contrast, the \ads mode suffers a potentially large cross feed from the favoured mode in the circumstance that both $D$ daughters are misidentified.
The invariant mass spectrum of such cross feed is broad but peaks around $m^{D^0}_{\rm PDG}$. 
It is reduced by vetoing any \ads candidate whose $D$ candidate mass under the exchange of its daughter track mass hypotheses, lies within $\pm15$~\mevcc of $m^{D^0}_{\rm PDG}$. 
Importantly for the measurements of $R_h^{\pm}$, this veto is also applied to the favoured mode.
With the $D$ mass selection and the $D$ daughter PID requirements, this veto reduces the rate of cross feed to an almost negligible rate of $(6\pm3)\times10^{-5}$.\\

Partially reconstructed events populate the invariant mass region below the \Bpm mass.
Such events may enter the signal region, especially where Cabibbo-favoured $B\to XD\pipm$ modes are misidentified as $B\to XD\Kpm$.
The large simulated sample of inclusive $\B_q\to DX$ decays, $q\in\{u,d,s\}$, is used to model this background.
After applying the selection, two non-parametric PDFs~\cite{Cranmer:2000du} are defined (for the $D\pipm$ and $D\Kpm$ selections) and used in the signal extraction fit.
These PDFs are applied to all four $D$ modes though two additional contributions are needed in specific cases.
In the $D\to\Kp\Km$ mode, $\Lambda_b^0\to [p^+\Km\pip]_{\Lambda_c}h^{-}$ enters if the pion is missed and the proton is reconstructed as a kaon.
In the $\Bpm\to D_{\!\ads}\Kpm$ mode, partially reconstructed $\Bsb\to\Dz\Kp\pim$ decays represent an important, Cabibbo-favoured background.
PDFs of both these sources are defined from simulation, smeared by the modest degradation in resolution observed in data.
When discussing these contributions, inclusion of the charge conjugate process is implied throughout.

\section{Signal yield determination}

\begin{figure*}[htb]
\begin{center}
\includegraphics[width=0.99\textwidth]{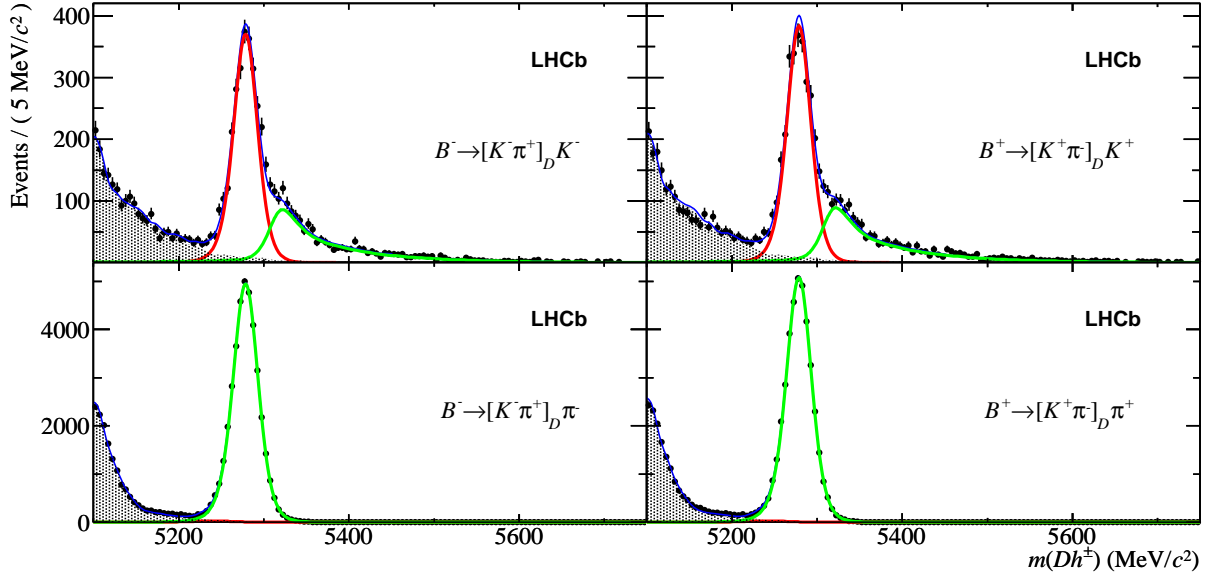}
\caption{Invariant mass distributions of selected $\Bpm\to[\Kpm\pi^{\mp}]_Dh^{\pm}$ candidates.
The left plots are \Bm candidates, \Bp are on the right. 
In the top plots, the bachelor track passes the ${\rm DLL}_{K\pi}>4$ cut and the $B$ candidates are reconstructed assigning this track the kaon mass. 
The remaining events are placed in the sample displayed on the bottom row and are reconstructed with a pion mass hypothesis.
The dark (red) curve represents the $\B\to D\Kpm$ events, the light (green) curve is $\B\to D\pipm$. 
The shaded contribution are partially reconstructed events and the total PDF includes the combinatorial component.
\label{fig:d2kpi}}
\end{center}
\end{figure*}

\begin{figure*}[htb]
\begin{center}
\includegraphics[width=0.99\textwidth]{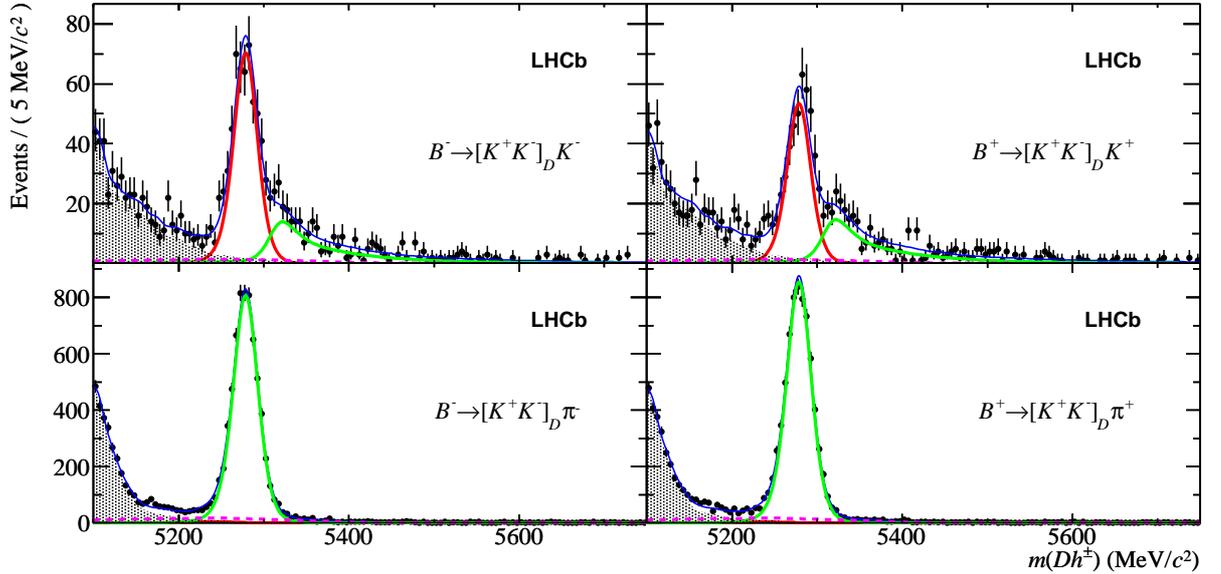}
\caption{Invariant mass distributions of selected $\Bpm\to[\Kp \Km]_Dh^{\pm}$ candidates. 
See the caption of Fig.~\ref{fig:d2kpi} for a full description.
The contribution from $\Lambda_b\to\Lambda_c^{\pm} h^{\mp}$ decays is indicated by the dashed line.
\label{fig:d2kk}}
\end{center}
\end{figure*}

\begin{figure*}[htb]
\begin{center}
\includegraphics[width=0.99\textwidth]{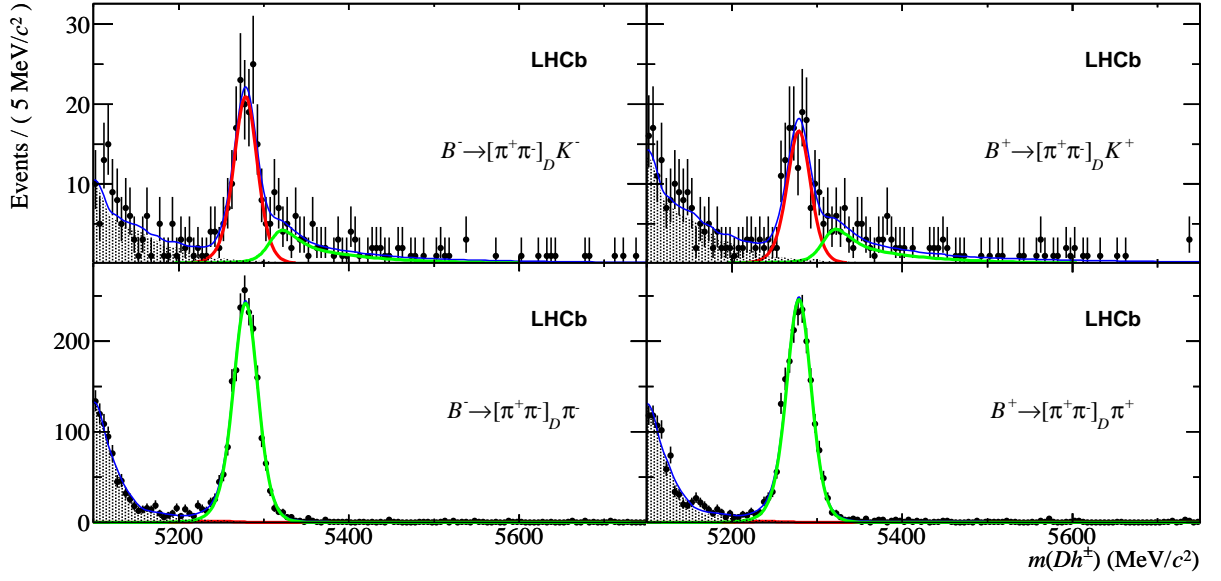}
\caption{Invariant mass distributions of selected $\Bpm\to[\pip\pim]_Dh^{\pm}$ candidates. 
See the caption of Fig.~\ref{fig:d2kpi} for a full description.
\label{fig:d2pipi}}
\end{center}
\end{figure*}

\begin{figure*}[htb]
\begin{center}
\includegraphics[width=0.99\textwidth]{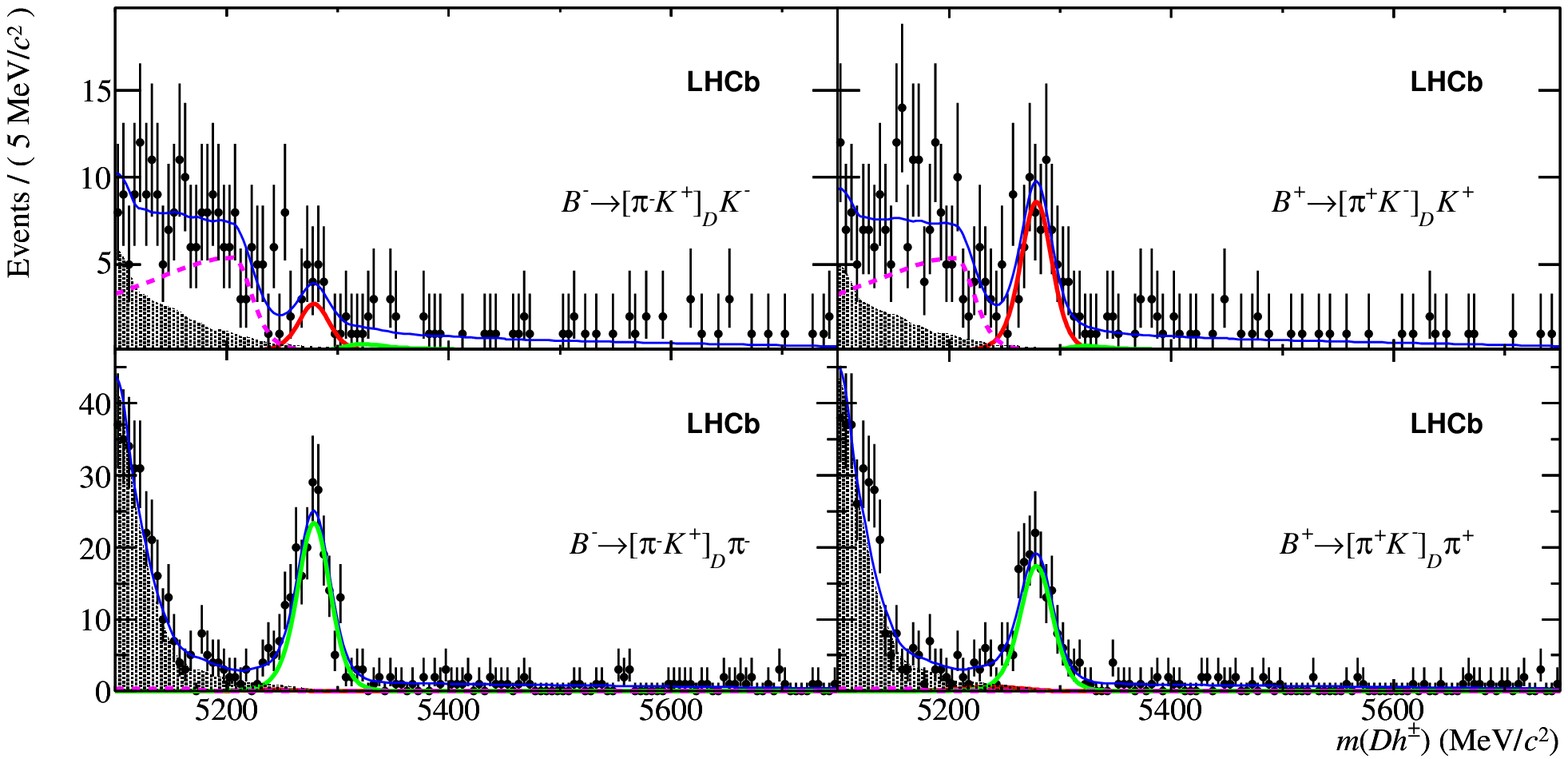}
\caption{Invariant mass distributions of selected $\Bpm\to[\pipm K^{\mp}]_Dh^{\pm}$ candidates. 
See the caption of Fig.~\ref{fig:d2kpi} for a full description.
The dashed line here represents the partially reconstructed, but Cabibbo favoured, $\Bs\to\Dzb \Km\pip$ and $\Bsb\to\Dz \Kp\pim$ decays where the pions are lost.
The pollution from favoured mode cross feed is drawn, but is too small to be seen.
\label{fig:d2pik}}
\end{center}
\end{figure*}

The observables of interest are determined with a binned maximum-likelihood fit to the invariant mass distributions of selected \B candidates~\cite{2003physics6116V}.
Sensitivity to \CP asymmetries is achieved by separating the candidates into  \Bm and \Bp samples.
$\Bpm\to D\Kpm$ events are distinguished from $\Bpm\to D\pipm$ using a PID cut on the ${\rm DLL}_{K\pi}$ of the bachelor track.
Events passing this cut are reconstructed as $D\Kpm$, events failing the cut are reconstructed as the $D\pipm$ final state.
The fit therefore comprises four subsamples --- $(\mathrm{plus , minus})\! \times\!(\mathrm{pass , fail})$ --- for each $D$ mode, fitted simultaneously and displayed in Figs.~\ref{fig:d2kpi}--\ref{fig:d2pik}.
The total PDF is built from four or five components representing the various sources of events in each subsample.

\begin{enumerate}
\item \underline{$\Bpm\to D\pipm$}.
In the sample failing the bachelor PID cut, a modified Gaussian function,
\begin{equation}
f(x) \propto \exp\left(\frac{-(x-\mu)^2}{2\sigma^2+(x-\mu)^2\alpha_{L,R}}\right)
\end{equation}
describes the asymmetric peak of mean $\mu$ and width $\sigma$ where $\alpha_L(x<\mu)$ and $\alpha_R(x>\mu)$ parameterise the tails.\\
True $\Bpm\to D\pipm$ events that pass the PID cut are reconstructed as $\Bpm\to D\Kpm$.
As these events have an incorrect mass assignment they form a displaced mass peak with a tail that extends to higher invariant mass.
These events are modelled by the sum of two Gaussian PDFs also altered to include tail components.
All parameters are allowed to vary except the lower-mass tail which is fixed to ensure fit stability and later considered amongst the systematic uncertainties.
These shapes are considered identical for \Bm and \Bp decays and for all four $D$ modes. This assumption is validated with simulation.
\item \underline{$\Bpm\to D\Kpm$}: 
In the sample that passes the ${\rm DLL}_{K\pi}$ cut on the bachelor, the same modified Gaussian function is used.
The mean and the two tail parameters are identical to those of the larger, $\Bpm\to D\pipm$ peak.
The width is $0.95\pm0.02$ times the $D\pipm$ width, as determined by a standalone study of the favoured mode.
Its applicability to the \CP modes is checked with simulation and a 1\% systematic uncertainty assigned.
Events failing the PID cut are described by a fixed shape that is obtained from simulation and later varied to assess the systematic error.
\item \underline{Partially reconstructed $\B\ra\D X$}: 
A fixed, non-parametric PDF, derived from simulation, is used for all subsamples.
The yield in each subsample varies independently, making no assumption of \CP symmetry.
\item \underline{Combinatoric background}: A linear approximation is adequate to describe the slope across the invariant mass spectrum considered.
A common parameter is used in all subsamples, though yields vary independently.
\item \underline{Mode-specific backgrounds}: 
In the $D\to KK$ mode, two extra components are used to model $\Lambda^0_b \to \Lambda_c^{+} h^{-}$ decays. 
Though the total contribution is allowed to vary, the shape and relative proportion of $\Lambda_c^{+} K^{-}$ and $\Lambda_c^{+} \pi^{-}$ are fixed.
This latter quantity is estimated at $0.060\pm0.015$, similar to the effective Cabibbo suppression observed in \B mesons.
For the $\Bpm\to D_{\!\ads}\Kpm$ mode, the shape of the $\Bsb\to\Dz\Kp\pim$ background is taken from simulation. 
In the fit, this yield is allowed to vary though the reported yield is consistent with the simulated expectation, as derived from the branching fraction~\cite{Aaij:2011tz} and the \bbbar hadronisation~\cite{Aaij:2011hi}.
\end{enumerate}

The proportion of $\Bpm\to Dh^{\pm}$ passing or failing the PID requirement is determined from a calibration analysis of a large sample of \Dstarpm decays reconstructed as $\Dstarpm\to D\pipm,\ D\to\Kmp\pipm$. 
In this calibration sample, the $K$ and $\pi$ tracks may be identified, with high purity, using only kinematic variables. 
This facilitates a measurement of the RICH-based PID efficiency as a function of track momentum, pseudorapidity and number of tracks in the detector.
By reweighting the calibration spectra in these variables to match the events in the $\Bpm\to D\pipm$ peak, the effective PID efficiency of the signal is deduced.
This data-driven technique finds a retention rate, for a cut of ${\rm DLL}_{K\pi}>4$ on the bachelor track, of 87.6\% and 3.8\% for kaons and pions, respectively.
A $1.0\%$ systematic uncertainty on the kaon efficiency is estimated from simulation. 
The $\Bpm\to D\pipm$ fit to data becomes visibly incorrect with variations to the fixed PID efficiency $>\pm0.2\%$ so this value is taken as the systematic uncertainly for pions.\\

A small negative asymmetry is expected in the detection of \Km and \Kp mesons due to their different interaction lengths.
A fixed value of $(-0.5\pm0.7)$\% is assigned for each occurrence of strangeness in the final state.
The equivalent asymmetry for pions is expected to be much smaller and ($0.0\pm0.7$)\% is assigned.
This uncertainty also accounts for the residual physical asymmetry between the left and right sides of the detector after summing both magnet-polarity datasets.
Simulation of \B meson production in $pp$ collisions suggests a small excess of \Bp over \Bm mesons. 
A production asymmetry of $(-0.8\pm0.7)$\% is assumed in the fit such that the combination of these estimates aligns with the observed raw asymmetry of $\Bpm\to\jpsi\Kpm$ decays at \lhcb~\cite{LHCb-PAPER-2011-024}. 
Ongoing studies of these instrumentation asymmetries will reduce the associated systematic uncertainty in future analyses.\\
The final $\Bpm\to Dh^{\pm}$ signal yields, after summing the events that pass and fail the bachelor PID cut, are shown in Table~\ref{tab:yields}.
The invariant mass spectra of all 16 $\Bpm\to[h^+ h^-]_Dh^{\pm}$ modes are shown in Figs.~\ref{fig:d2kpi}--\ref{fig:d2pik}. 
Regarding the $\Bpm\to D\pipm$ mass resolution: respectively, $14.1\pm0.1$, $14.2\pm0.1$ and $14.2\pm0.2$~\mevcc are found for the $D\to KK$, $K\pi$ and $\pi\pi$ modes with common tail parameters
$\alpha_L=0.115\pm0.003$ and $\alpha_R=0.083\pm0.002$. As explained above, the $\Bpm\to D\Kpm$ widths are fixed relative to these values.\\

\vspace{-0.5cm}
\begin{table}[!htp]
\begin{center}
\begin{small}
\caption{Corrected event yields. 
\label{tab:yields}}
\begin{tabular}{cl|cc}
%\hline\hline
\multicolumn{1}{l}{\Bpm mode} & $D$ mode            & \Bm & \Bp \\
\hline&&\\[-2.5ex]
\multirow{4}{*}{$D\Kpm$} & $\Kpm\pimp$  & $\phantom{0}3170\pm\phantom{0}83$ & $\phantom{0}3142\pm\phantom{0}83$ \\
                         & $\Kpm\Kmp$   & $\phantom{00}592\pm\phantom{0}40$ & $\phantom{00}439\pm\phantom{0}30$ \\
                         & $\pipm\pimp$ & $\phantom{00}180\pm\phantom{0}22$ & $\phantom{00}137\pm\phantom{0}16$ \\
                         & $\pipm\Kmp$  & $\phantom{000}23\pm\phantom{00}7$ & $\phantom{000}73\pm\phantom{0}11$ \\
\hline&&\\[-2.5ex]
\multirow{4}{*}{$D\pipm$}& $\Kpm\pimp$  &           $40767\pm310$           &           $40774\pm310$           \\
                         & $\Kpm\Kmp$   & $\phantom{0}6539\pm129$           & $\phantom{0}6804\pm135$           \\
                         & $\pipm\pimp$ & $\phantom{0}1969\pm\phantom{0}69$ & $\phantom{0}1973\pm\phantom{0}69$ \\
                         & $\pipm\Kmp$  & $\phantom{00}191\pm\phantom{0}16$ & $\phantom{00}143\pm\phantom{0}14$ \\
%\hline\hline
\end{tabular}
\end{small}
\end{center}
\end{table}

The ratio of partial widths relates to the ratio of event yields by the relative efficiency with which $\Bpm\to\Dz\Kpm$ and $\Bpm\to\Dz\pipm$ decays are reconstructed.
This ratio, estimated from simulation, is 1.012, 1.009 and 1.005 for $D\to KK,K\pi,\pi\pi$ respectively.
A 1.1\% systematic uncertainty, based on the finite size of the simulated sample, accounts for the imperfect modelling of the relative pion and kaon absorption in the tracking material.\\

The fit is constructed such that the observables of interest are parameters of the fit and all systematic uncertainties discussed above enter the fit as constant numbers in the model.
To evaluate the effect of these systematic uncertainties, the fit is rerun many times varying each of the systematic constants by its uncertainty.
The resulting spread (RMS) in the value of each observable is taken as the systematic uncertainty on that quantity and is summarised in Table~\ref{tab:syst}.
Correlations between the uncertainties are considered negligible so the total systematic uncertainty is just the sum in quadrature.
For the ratios of partial widths in the favoured and \CP modes, the uncertainties on the PID efficiency and the relative width of the $D\Kpm$ and $D\pipm$ peaks dominate.
These sources also contribute in the \ads modes, though the assumed shape of the $\Bsb\to\Dz\Kp\pim$ background is the largest source of systematic uncertainty in the $\Bpm\to D_{\!\ads}\Kpm$ case.
For the \CP asymmetries, instrumentation asymmetries at \lhcb are the largest source of uncertainty.
\begin{table}[!htp]
\begin{center}
\begin{small}
\caption{Systematic uncertainties on the observables.
PID refers to the fixed efficiency of the ${\rm DLL}_{K\pi}$ cut on the bachelor track.
PDFs refers to the variations of the fixed shapes in the fit.
``Sim" refers to the use of simulation to estimate relative efficiencies of the signal modes which includes the branching fraction estimates of the $\Lambda_b^0$ background.
$A_{\rm instr.}$ quantifies the uncertainty on the production, interaction and detection asymmetries.
\label{tab:syst}}
\begin{tabular}{lcccc|c}
$\times 10^{-3}$  &  PID & PDFs & Sim &  $A_{\rm instr.}$ & Total\\
\hline 
$R_{K/\pi}^{K\pi}$  & 1.4  & 0.9  & 0.8  & 0  & 1.8\\
$R_{K/\pi}^{KK}$  & 1.3  & 0.8  & 0.9  & 0  & 1.8\\
$R_{K/\pi}^{\pi\pi}$  & 1.3  & 0.6  & 0.8  & 0  & 1.7\\
$A_{\pi}^{K\pi}$  & 0  & 1.0  & 0  & 9.4  & 9.5\\
$A_{K}^{K\pi}$  & 0.2  & 4.1  & 0  & 16.9  & 17.4\\
$A_{K}^{KK}$  & 1.6  & 1.3  & 0.5  & 9.5  & 9.7\\
$A_{K}^{\pi\pi}$  & 1.9  & 2.3  & 0  & 9.0  & 9.5\\
$A_{\pi}^{KK}$  & 0.1  & 6.6  & 0  & 9.5  & 11.6\\
$A_{\pi}^{\pi\pi}$  & 0.1  & 0.4  & 0  & 9.9  & 9.9\\
$R_{K}^-$  & 0.2  & 0.4  & 0  & 0.1  & 0.4\\
$R_{K}^+$  & 0.4  & 0.5  & 0  & 0.1  & 0.7\\
$R_{\pi}^-$  & 0.01  & 0.03 & 0  & 0.07  & 0.08\\
$R_{\pi}^+$  & 0.01  & 0.03  & 0  & 0.07  & 0.07\\
\end{tabular}
\end{small}
\end{center}
\end{table}

\section{Results}
The results of the fit with their statistical uncertainties and assigned systematic uncertainties are:
\begin{eqnarray*}
R_{K/\pi}^{K\pi}&=& \phantom{-} 0.0774 \pm 0.0012 \pm 0.0018 \\
R_{K/\pi}^{KK} &=& \phantom{-} 0.0773 \pm 0.0030 \pm 0.0018 \\
R_{K/\pi}^{\pi\pi}&=& \phantom{-} 0.0803 \pm 0.0056 \pm 0.0017 \\
A_{\pi}^{K\pi} &=& -0.0001 \pm 0.0036 \pm 0.0095 \\
A_{K}^{K\pi} &=&  \phantom{-} 0.0044 \pm 0.0144 \pm 0.0174 \\
A_{K}^{KK} &=& \phantom{-} 0.148 \pm 0.037 \pm 0.010 \\
A_{K}^{\pi\pi}  &=& \phantom{-} 0.135 \pm 0.066 \pm 0.010 \\
A_{\pi}^{KK}  &=& -0.020 \pm 0.009 \pm 0.012 \\
A_{\pi}^{\pi\pi}&=& -0.001 \pm 0.017 \pm 0.010 \\
R_{K}^- &=& \phantom{-} 0.0073 \pm 0.0023 \pm 0.0004 \\
R_{K}^+ &=& \phantom{-} 0.0232 \pm 0.0034 \pm 0.0007 \\
R_{\pi}^-&=& \phantom{-} 0.00469 \pm 0.00038 \pm 0.00008 \\
R_{\pi}^+ &=& \phantom{-} 0.00352 \pm 0.00033 \pm 0.00007.
\end{eqnarray*}

\noindent
From these measurements, the following quantities can be deduced:
\begin{eqnarray*}
R_{\CP+} &\approx& <R_{K/\pi}^{KK} ,R_{K/\pi}^{\pi\pi} > / R_{K/\pi}^{K\pi} \\&=& \phantom{-} 1.007 \pm 0.038 \pm 0.012 \\
A_{\CP+} &=& <A_{K}^{KK} ,A_{K}^{\pi\pi} >\\&=&\phantom{-}  0.145 \pm 0.032 \pm 0.010 \\
R_{\ads(K)} &=& (R_{K}^- + R_{K}^+) / 2 \\&=&\phantom{-}  0.0152 \pm 0.0020 \pm 0.0004 \\
A_{\ads(K)} &=&  (R_{K}^- - R_{K}^+) / (R_{K}^- + R_{K}^+) \\ &=& -0.52 \pm 0.15 \pm 0.02 \\
R_{\ads(\pi)} &=& (R_{\pi}^- + R_{\pi}^+) / 2 \\&=&\phantom{-}  0.00410 \pm 0.00025 \pm 0.00005 \\
A_{\ads(\pi)} &=& (R_{\pi}^- - R_{\pi}^+) / (R_{\pi}^- + R_{\pi}^+) \\ &=& \phantom{-} 0.143 \pm 0.062 \pm 0.011,
\end{eqnarray*}
where the correlations between systematic uncertainties are taken into account in the combination and angled brackets indicate weighted averages.
The above definition of $R_{\CP+}$ is only approximate and is used for experimental convenience. It assumes the absence of \CP violation in $\Bpm\to\D\pipm$ and the favoured $\Bpm\to\D\Kpm$ modes.
The exact definition of $R_{\CP+}$ is 
\begin{equation}
\frac{\Gamma(\Bm\to D_{\CP+}\Km)+\Gamma(\Bp\to D_{\CP+}\Kp)}{\Gamma(\Bm\to \Dz\Km)}
\end{equation}
so an additional, and dominant, 1\% systematic uncertainty accounts for the approximation.
For the same reason, a small addition to the systematic uncertainty of $R_{K/\pi}^{K\pi}$ is needed to quote this result as the ratio of \Bpm branching fractions,
\begin{equation*}
\frac{\mathcal{B}(\Bm\to \Dz\Km)}{\mathcal{B}(\Bm\to \Dz\pim)} = (7.74 \pm 0.12 \pm 0.19)\%. \\
\end{equation*}
To summarise, the $\Bpm\to D\Kpm$ \ads mode is observed with \mbox{$\sim10\sigma$} statistical significance when comparing the maximum likelihood to that of the null hypothesis.
This mode displays evidence ($4.0\sigma$) of a large negative asymmetry, consistent with the asymmetries reported by previous experiments~\cite{Belle:2011ac,delAmoSanchez:2010dz,Aaltonen:2011uu}.
The $\Bpm\to D\pipm$ \ads mode shows a hint of a positive asymmetry with $2.4\sigma$ significance.
The $KK$ and $\pi\pi$ modes both show positive asymmetries. 
The statistical significance of the combined asymmetry, $A_{\CP+}, $ is $4.5\sigma$ which is similar to that reported in~\cite{delAmoSanchez:2010ji,Aaltonen:2009hz} albeit with a smaller central value.
All these results contain dependence on the weak phase \g and will form an important contribution to a future measurement of this parameter.\\

Assuming the \CP-violating effects in the \CP and \ads modes are due to the same phenomenon (namely the interference of $b\to c\bar{u}s$ and $b\to u\bar{c}s$ transitions) 
we compare the maximum likelihood with that under the null-hypothesis in all three $D$ final states where the bachelor is a kaon.
This log-likelihood difference is diluted by the non-negligible systematic uncertainties in $A_{\CP+}$ and $A_{\ads(K)}$ which are dominated by the instrumentation asymmetries and hence are highly correlated.
In conclusion, with a total significance of $5.8\sigma$, direct \CP violation in $\Bpm\to D\Kpm$ decays is observed.

\section*{Acknowledgements}
\noindent 
We express our gratitude to our colleagues in the CERN accelerator
departments for the excellent performance of the LHC. We thank the
technical and administrative staff at CERN and at the LHCb institutes,
and acknowledge support from the National Agencies: CAPES, CNPq,
FAPERJ and FINEP (Brazil); CERN; NSFC (China); CNRS/IN2P3 (France);
BMBF, DFG, HGF and MPG (Germany); SFI (Ireland); INFN (Italy); FOM and
NWO (The Netherlands); SCSR (Poland); ANCS (Romania); MinES of Russia and
Rosatom (Russia); MICINN, XuntaGal and GENCAT (Spain); SNSF and SER
(Switzerland); NAS Ukraine (Ukraine); STFC (United Kingdom); NSF (USA). 
We also acknowledge the support received from the ERC under FP7 and the Region Auvergne.

\clearpage
\onecolumn

\ifx\mcitethebibliography\mciteundefinedmacro
\PackageError{LHCb.bst}{mciteplus.sty has not been loaded}
{This bibstyle requires the use of the mciteplus package.}\fi
\providecommand{\href}[2]{#2}

\end{document}